\documentclass[sigconf,screen,authordraft,review=false,timestamp=false]{acmart}

\AtBeginDocument{%
  }

\copyrightyear{2024}
\acmYear{2024}
\setcopyright{rightsretained}
\acmConference[Websci Companion '24]{16th ACM Web Science Conference}{May 21--24, 2024}{Stuttgart, Germany}
\acmBooktitle{16th ACM Web Science Conference (Websci Companion '24), May 21--24, 2024, Stuttgart, Germany}
\acmDOI{10.1145/3630744.3658414}
\acmISBN{979-8-4007-0453-6/24/05}

\usepackage[font=footnotesize,skip=0pt]{caption}

\begin{document}

\title{Amplifying Academic Research through YouTube: Engagement Metrics as Predictors of Citation Impact}

\author{Olga Zagovora}
\email{olga.zagovora@rptu.de}
\orcid{0000-0002-4693-9668}
\affiliation{%
  \institution{RPTU Kaiserslautern-Landau}
  \city{Landau}
  \country{Germany}}
\affiliation{%
  \institution{DFKI GmbH}
  \city{Kaiserslautern}
  \country{Germany}}

\author{Talisa Schwall}
\email{talisa.schwall@rptu.de}
\orcid{0009-0008-1550-7599}
\affiliation{
  \institution{RPTU Kaiserslautern-Landau}
  \city{Landau}
  \country{Germany}}

\author{Katrin Weller}
\email{katrin.weller@gesis.org}
\orcid{0000-0003-3799-1146}
\affiliation{
  \institution{GESIS - Leibniz Institute for the Social Sciences}
  \city{Cologne}
  \country{Germany}}

\begin{abstract}
\vspace{-2mm}
This study explores the interplay between YouTube engagement metrics and the academic impact of cited publications within video descriptions, amid declining trust in traditional journalism and increased reliance on social media for information. By analyzing data from Altmetric.com and YouTube's API, it assesses how YouTube video features relate to citation impact. Initial results suggest that videos citing scientific publications and garnering high engagement—likes, comments, and references to other publications—may function as a filtering mechanism or even as a predictor of impactful research. 
\end{abstract}




\maketitle

\vspace{-2mm}
\section{Introduction}
\vspace{-2mm}

Amid declining trust in traditional journalism, people increasingly turn to the internet and social media for information\cite{newman_digital_2023}, a shift particularly emphasized among those doubting mainstream media. This shift is crucial in a "permacrisis" context, making the analysis of academic source usage on social media platforms like YouTube vital. 

This study focuses on YouTube's role in disseminating scientific knowledge to its vast audience, over 5.6 billion monthly visitors (\url{https://www.statista.com/statistics/1201889/most-visited-websites-worldwide-unique-visits/})
, positioning it as a significant source to a wide public for science, health, and medicine information\cite{velho_profiles_2020}. Study \cite{velho_profiles_2020} underscores science YouTubers' primary objectives to enhance the public's scientific understanding and combat misinformation. Initial findings in \cite{zagovora_youtube_2018} suggest a slight positive link between the citation impact of referenced publications and YouTube comment volume, hinting at comments as valuable for altmetrics studies. This leads to investigating if YouTube's engagement metrics could reflect a video's and its cited publications' influence. Moreover, comments citing additional publications might mirror academic discussions, akin to traditional citation impact measures \cite{striewski_scientific_2022}. With this in mind, we aim to explore: 
\begin{itemize}
\vspace{-1.6mm}
\item {\bfseries RQ1:} Is there a correlation between the number of comments with links to other publications on YouTube videos and the citation impact of the papers referenced in these videos?
\item {\bfseries RQ2:} Which YouTube video attributes are linked to heightened research impact?
\end{itemize}

\vspace{-7mm}
\section{Data Collection \& Method}
\vspace{-2mm}
This study collected initial data from \url{https://www.altmetric.com/}, a tool that aggregates altmetrics by compiling mentions of research outputs from social media and mainstream media. We focused on YouTube videos that referenced scholarly publications in their descriptions, resulting in a dataset of 41,445 videos (posted 2006-2017). Additional data on video characteristics and audience engagement were gathered using Google's YouTube Data API v3.

\begin{figure}
\begin{center}
  \includegraphics[width=1.05\columnwidth]{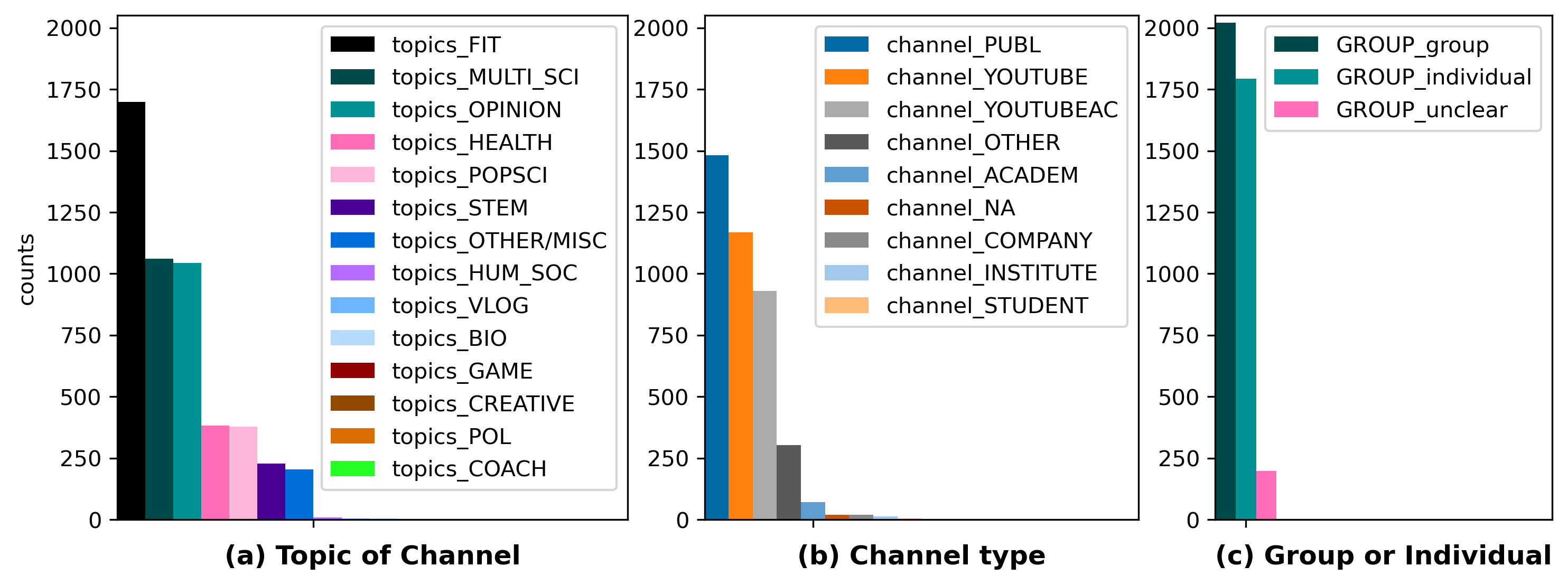}
  \centering
  \caption{Counts of (a) topic of channel\textmd{: FIT - fitness and nutrition, MULTI\_SCI - multidisciplinary channels with professional science perspective, OPINION - personal opinion and alternative news, HEALTH - medicine and health sciences, POPSCI - multidisciplinary channels with popular science perspective, STEM - STEM fields of science, OTHER/MISC - other topics or miscellaneous, HUM\_SOC - humanities and social sciences, VLOG - video blogs, BIO - biology and life sciences, GAME, CREATIVE - creativity, POL - politics and activism, COACH - coaching;} (b) channel type\textmd{: PUBL - publishers and journals, YOUTUBE - (semi-) professional YouTube users who have an established online presence on the platform, YOUTUBEAC - those who qualify into the previous group but also hold an academic degree in the field of their channel topic, OTHER - those who do not fit in any of the categories, ACADEM - academics or groups of academics, NA - no available information, COMPANY, INSTITUTE - research institutes or organizations, STUDENT - student accounts;} (c) channels moderated by a group, an individual\textmd{, or when it is} unclear.}
  \Description{Three plots that describe counts of different categories of the dataset. }
  \label{fig:topics}
\end{center}
\vspace{-2mm}
\end{figure}

We used a manually annotated dataset \cite{zagovora_between_nodate} that labels YouTube channels by topic, management (individual or group), and whether they possess academic backgrounds, represent research institutions, or hold academic degrees (Figure~\ref{fig:topics}). This dataset was combined with another containing comments referencing publications \cite{striewski_scientific_2022}, allowing us to count publication mentions in video comments. This integration resulted in 4,012 mentions of 2,691 publications across 2,202 videos.

Citation numbers for publications were sourced from Altmetric.com, normalized based on publication year and Scopus category, and log-transformed \cite{thelwall_three_2017} to mitigate data skewness and adjust for citation accumulation over time. These adjusted values are further referred to as the publication's citation impact.

To address our RQs, we employed a regression model using the following as independent variables:
the number of videos referencing a publication, the like-to-dislike ratios, 
the count of comments mentioning other publications, and - based on the manual categorization described above - 
the channel characteristics (type, topic, and moderation by an individual or group). The dependent variable was the publication's citation impact, 
with the reference groups being {\itshape channel INSTITUTE}, {\itshape topics BIO}, {\itshape GROUP group}. To avoid multicollinearity, we excluded video views and comment counts due to their correlation with 
the number of comments mentioning other publications. 
Log transformations were applied to: the comment counts mentioning other publications, the number of videos referencing the same publication, and the like-to-dislike ratios. 

\vspace{-4mm}
\section{Results \& Discussion} 
\vspace{-2mm}
The preliminary findings from the linear regression analysis (Table~\ref{tab:table1}) suggest a meaningful relationship between the online engagement metrics of videos on YouTube and the academic impact of the publications referenced within these videos. Specifically, the analysis found positive correlations with the citation impact for three key metrics: the number of videos referencing publications, the ratio of likes to dislikes on videos, and the number of comments containing references to other publications. The positive correlation indicates a sort of selective amplification process. Publications mentioned in videos that garner attention in the form of likes and active discussion in comments are likely being selectively chosen for their relevance or quality. This selection process by content creators and the subsequent engagement by viewers may serve as an "informal peer review", signaling the value and impact of the research. The findings suggest that social media, particularly YouTube in this context, acts as a filter that potentially can highlight the visibility of impactful research.

\begin{table}
\caption{OLS Regression Results\textmd{: The number of videos referencing publications, {\itshape videos count log}, the ratio of likes to dislikes on videos, {\itshape LikeToDislike log}, and the number of comments containing references to other publications, {\itshape commentToPub log}, are positively correlated with the citation impact of referenced papers.}}

\label{tab:table1}
\begin{tabular*}{0.82\columnwidth}{c|}
\toprule
\end{tabular*}

\resizebox{0.75\columnwidth}{!}{
\begin{tabular}{lclc}
\textbf{Dep. Variable:}         &  citation impact  & \textbf{  R-squared:         } &     0.158   \\
\textbf{Model:}                 &       OLS        & \textbf{  Adj. R-squared:    } &     0.152   \\
\textbf{Method:}                &  Least Squares   & \textbf{  F-statistic:       } &     28.69   \\
\textbf{Date:}                  & Thu, 15 Feb 2024 & \textbf{  Prob (F-statistic):} & 7.09e-128   \\
\textbf{Time:}                  &     12:45:30     & \textbf{  Log-Likelihood:    } &   -1282.6   \\
\textbf{No. Observations:}      &        4012      & \textbf{  AIC:               } &     2619.   \\
\textbf{Df Residuals:}          &        3985      & \textbf{  BIC:               } &     2789.   \\
\textbf{Df Model:}              &          26      & \textbf{                     } &             \\
\end{tabular}
}
\begin{tabular*}{0.82\columnwidth}{c|}
\toprule
\end{tabular*}

\resizebox{0.82\columnwidth}{!}{
\begin{tabular}{lcccccc}
                                & \textbf{coef} & \textbf{std err} & \textbf{t} & \textbf{P$> |$t$|$} & \textbf{[0.025} & \textbf{0.975]}  \\
\midrule
\textbf{const}                  &       0.6228  &        0.138     &     4.508  &         0.000        &        0.352    &        0.894     \\
\textbf{videos count log}     &       0.0274  &        0.009     &     2.886  &         0.004        &        0.009    &        0.046     \\
\textbf{LikeToDislike log}     &       0.0357  &        0.008     &     4.751  &         0.000        &        0.021    &        0.050     \\
\textbf{commentToPub log} &       0.0170  &        0.007     &     2.365  &         0.018        &        0.003    &        0.031     \\
\textbf{channel ACADEM}        &      -0.1308  &        0.114     &    -1.149  &         0.251        &       -0.354    &        0.093     \\
\textbf{channel COMPANY}       &       0.0883  &        0.157     &     0.562  &         0.574        &       -0.220    &        0.396     \\
\textbf{channel NA}            &      -0.5015  &        0.127     &    -3.954  &         0.000        &       -0.750    &       -0.253     \\
\textbf{channel OTHER}         &      -0.0982  &        0.099     &    -0.988  &         0.323        &       -0.293    &        0.097     \\
\textbf{channel PUBL}          &      -0.1486  &        0.095     &    -1.564  &         0.118        &       -0.335    &        0.038     \\
\textbf{channel STUDENT}       &      -0.0711  &        0.211     &    -0.337  &         0.736        &       -0.485    &        0.343     \\
\textbf{channel YOUTUBE}       &       0.1493  &        0.140     &     1.066  &         0.287        &       -0.125    &        0.424     \\
\textbf{channel YOUTUBEAC}     &       0.0924  &        0.137     &     0.672  &         0.501        &       -0.177    &        0.362     \\
\textbf{topics COACH}          &       0.2423  &        0.348     &     0.697  &         0.486        &       -0.440    &        0.924     \\
\textbf{topics CREATIVE}       &      -0.3418  &        0.206     &    -1.658  &         0.097        &       -0.746    &        0.062     \\
\textbf{topics FIT}            &       0.0428  &        0.061     &     0.700  &         0.484        &       -0.077    &        0.163     \\
\textbf{topics GAME}           &      -0.2032  &        0.206     &    -0.987  &         0.324        &       -0.607    &        0.200     \\
\textbf{topics HEALTH}         &       0.0082  &        0.097     &     0.085  &         0.932        &       -0.182    &        0.199     \\
\textbf{topics HUM\_SOC}       &      -0.2683  &        0.137     &    -1.956  &         0.051        &       -0.537    &        0.001     \\
\textbf{topics MULTI\_SCI}     &       0.0615  &        0.096     &     0.638  &         0.523        &       -0.127    &        0.250     \\
\textbf{topics OPINION}        &      -0.0056  &        0.034     &    -0.165  &         0.869        &       -0.072    &        0.060     \\
\textbf{topics OTHER/MISC}     &       0.1726  &        0.098     &     1.769  &         0.077        &       -0.019    &        0.364     \\
\textbf{topics POPSCI}         &      -0.0274  &        0.095     &    -0.289  &         0.772        &       -0.213    &        0.158     \\
\textbf{topics STEM}           &       0.1181  &        0.094     &     1.254  &         0.210        &       -0.067    &        0.303     \\
\textbf{topics VLOG}           &       0.1330  &        0.169     &     0.786  &         0.432        &       -0.199    &        0.465     \\
\textbf{topics POL}            &       0.3520  &        0.351     &     1.003  &         0.316        &       -0.336    &        1.040     \\
\textbf{GROUP individual}      &      -0.0726  &        0.076     &    -0.961  &         0.337        &       -0.221    &        0.075     \\
\textbf{GROUP unclear}         &      -0.0062  &        0.114     &    -0.054  &         0.957        &       -0.230    &        0.217     \\
\bottomrule
\end{tabular}
}
\resizebox{0.6\columnwidth}{!}{
\begin{tabular}{lclc}
\textbf{Omnibus:}       &  4.866 & \textbf{  Durbin-Watson:     } &    1.156  \\
\textbf{Prob(Omnibus):} &  0.088 & \textbf{  Jarque-Bera (JB):  } &    5.253  \\
\textbf{Skew:}          & -0.034 & \textbf{  Prob(JB):          } &   0.0723  \\
\textbf{Kurtosis:}      &  3.164 & \textbf{  Cond. No.          } &     210.  \\
\bottomrule
\end{tabular}
}
\end{table}

Moreover, the correlation with citation impact underscores the role of social media in extending the reach and influence of academic research beyond traditional academic circles. Videos that reference academic publications and generate significant engagement can bridge the gap between academia and the broader public, potentially influencing both academic citations and public discourse.
 
At the same time, we have observed that channels categorized as {\itshape NA}, that is, channels with no clear topic, no information on the purpose of the channel, and no details about the people involved in creating content, tend to reference publications with low impact. For researchers in the field of scientometrics, this suggests that not every reference to a publication should be treated equally; instead, certain credibility metrics should be incorporated into the filtering of channels if these altmetrics are to be used for approximating citation impact.

Our study focuses on a dataset of videos from 2006 to 2017, which may limit the generalizability of its findings to current trends in YouTube engagement and academic citation practices. One might argue that socially relevant topics (e.g., COVID) could further skew the interest in medical topics. Nevertheless, our previous studies \cite{zagovora_youtube_2018, zagovora_between_nodate} have shown that Health\&Medicine were among the most popular and most extensively covered topics even before the pandemic. 

This analysis opens avenues for further investigation into the mechanisms through which online engagement relates to academic impact. Future research will incorporate all three groups of features associated with social media entities: features of actors who create social media content, user engagement with social media content, and features of the social media entities themselves. The third group of features might include the topic of the video, the length of the video, and the delay between the publication date of the video and the research output. 
\vspace{-3mm}
\begin{acks}
\vspace{-2mm}
We thank Altmetric.com for providing data for research purposes.
\end{acks}
\vspace{-2mm}

\bibliographystyle{ACM-Reference-Format}
\vspace{-2mm}
\bibliography{websci24}

\end{document}